\documentclass[12pt,preprint]{article}
\usepackage{amssymb}
\usepackage{amsmath}
\usepackage{graphics}
\usepackage{epsfig}

\renewcommand{\vec}[1]{{\bf #1}}
\newcommand{\eqb}{\begin{equation}}
\newcommand{\eqe}{\end{equation}}
\newcommand{\dmb}{\begin{displaymath}}
\newcommand{\dme}{\end{displaymath}}
\newcommand{\pd}{\partial}

\newcommand{\eab}{\begin{eqnarray}}
\newcommand{\eae}{\end{eqnarray}}
\newcommand{\ra}{\right\rangle}
\newcommand{\la}{\left\langle}

\newcommand{\be}{\begin{equation}}
\newcommand{\ee}{\end{equation}}

\input{tcilatex}

\begin{document}

\begin{titlepage}
\begin{flushright} 
\end{flushright}
\vspace{0.6cm}

\begin{center}
\Large{CMB dipole revisited}
\vspace{1.5cm}

\large{Josef Ludescher and Ralf Hofmann}

\end{center}
\vspace{1.5cm} 

\begin{center}
{\em Institut f\"ur Theoretische Physik\\ 
Universit\"at Heidelberg\\ 
Philosophenweg 16\\ 
69120 Heidelberg, Germany}
\end{center}
\vspace{1.5cm}

\begin{abstract}

We re-evaluate the build-up of a horizon-sized temperature 
profile of amplitude $\delta T/T \sim 10^{-3}$ at $z\sim 1$ in light of an improved determination of the 
black-body anomaly, based on a pure SU(2) Yang-Mills theory, 
and the correction of a mistake in deriving the evolution equation for $\delta T$. 
Our present results for the temperature profiles hardly are distinguishable 
from those published previously.      
   
\end{abstract} 

\end{titlepage}

\section{Introduction}

Judged by Standard-Model physics (photon propagation based on a U(1) gauge group, cosmologically 
scale invariant primordial perturbation spectrum) the observation that the  
dipole-subtracted CMB TT correlation function $C(\theta)$ is consistent with zero for 
angular separation of 60 degrees and larger and that the low-lying 
CMB multipoles ($l=2,3$) seem to be statistically aligned are incomprehensible 
\cite{largeangles}. This appears to be in line with recent radio-source investigations 
\cite{RudnickBrownWilliams} in the neighbourhood of the CMB cold spot (on 4 and 10 degrees 
resolution one observes an amplitude of -73\,$\mu$K and -20\,$\mu$K, respectively) which suggest 
that this cold spot is not of any primordial 
origin. Rather, it seems to be 
related to physics operating at a redshift of 
about $z\sim 1$. When invoking the late integrated Sachs-Wolfe effect this physics seems to have 
created a region void of matter of $\sim 140\,$Mpc radius. 
Notice, however, that there may be an alternative explanation, which does not relate to the spatial distribution of 
gravitational potentials but to the occurrence of a low-temperature, low-frequency anomaly in black-body 
spectra generated by the nonabelian effects of an SU(2) gauge symmetry underlying the {\sl propagation} of 
thermalized photons \cite{SzopaHofmann2007}. It is interesting though  
that the late integrated Sachs-Wolfe effect points to an anomaly which is 
operative at the same value of redshift $z\sim 1$ where the 
black-body anomaly is maximal.        

The approach of \cite{SzopaHofmann2007} is to treat the temperature fluctuation 
$\delta T$ as a scalar field whose evolution is 
subject to primordial, Gaussian initial conditions at sufficiently high redshift 
$z$. Doing justice to luminosity-redshift curves extracted from calibrated 
supernovae and to best-fitted large-$l$ CMB data, a standard $\Lambda$CDM background 
cosmology was used in \cite{SzopaHofmann2007} to drive the Universe's expansion. In this given cosmological background, the evolution 
of $\delta T$ is, according to \cite{SzopaHofmann2007}, in addition 
determined by coefficient functions provided by the modified black-body 
spectrum. These functions enter into an evolution equation derivable from an action principle with an 
 assumption about the relative normalization of kinetic 
versus potential term in the action being made. So far, this 
normalization is determined empirically although it should, in principle, 
be derivable from the SU(2) gauge theory of scale $\Lambda_{\tiny\mbox{CMB}}\sim 10^{-4}$\,eV postulated to underly 
photon propagation \cite{Hofmann2005,HofJHW2005,SHG2006I,SHG2006II,SHSG2007,LH2008}.     

As a consequence, the rapid build-up of a horizon-sized $\delta T$ profile of 
CMB dipole strength is predicted to occur at $z\sim 1$. In general, 
the dipole subtracted angular TT correlation function $C(\theta)$ is obtained by integrating the map of 
fluctuations $\delta T$ 
at $z=0$ along two lines of sight which are 
separated by the angle $\theta$ and by subsequently averaging over maps subject to varying 
initial conditions and over pairs of lines of sight\footnote{These maps are dipole subtracted.}.  

The purpose of the present work is to investigate to what extent the results in 
\cite{SzopaHofmann2007} are modified when subjecting the $z$ evolution of the $\delta T$ 
maps to the exact black-body anomaly computed in \cite{LH2008} and 
to a corrected evolution equation for $\delta T$. Notice that in \cite{SzopaHofmann2007} a black-body anomaly was computed 
based on the approximation that the photon is on its naive mass shell ($p^2=0$) \cite{SHG2006I}. Compared 
with the selfconsistent determination of the screening function $G$ \cite{LH2008} 
this yields good results for $T\le 2\,T_c$ where $T_c$ denotes the critical temperature 
for the deconfining-preconfining phase transition. However, for $T>2\,T_c$ the exact (selfconsistently determined)
result for $G$ falls off much faster with increasing $T$ and the spectral gap 
closes more rapidly than the results obtained with $p^2=0$ seemed to suggest. 

The paper is organized as follows. In Sec.\,\ref{Spher} we briefly discuss the 
objects that are essential to our present investigation, namely, the 
accurate determination of the spectral black-body anomaly and the cosmological 
evolution equation for the temperature fluctuation $\delta T$. In Sec.\,\ref{OVN} 
we compare the old results, obtained with the erroneous evolution equation 
and an inaccurate black-body anomaly, with the ones obtained by rectifying 
these problems observing that on the level of spatial 
spherical symmetry practically no difference occurs. 
Finally, we give a short summary in Sec.\,\ref{SC}.

\section{Black-body anomaly 
and cosmological evolution of a spherical temperature profile\label{Spher}}

In \cite{SHG2006I} the polarization tensor of the massless mode was 
computed to one-loop accuracy in deconfining SU(2) Yang-Mills thermodynamics under the 
assumption that the external four-momentum $p$ is on the 
free mass shell: $p^2=0$. To address the effects determining the photon dispersion 
law fully\footnote{It was shown in \cite{KH2007} that three-loop contributions 
to the pressure of deconfining SU(2) Yang-Mills thermodynamics 
are suppressed as compared to the two-loop correction by a factor of $10^{-3}$ or smaller. 
This result relates to the photon polarization by cutting the 
massless lines in the corresponding  pressure corrections. Thus 
practically no modification of the one-loop result for the 
photon polarization arises from higher loops.} the modified on-shell dispersion relation, as it 
arises from the resummed one-loop polarization, needs to be taken into account in a selfconsistent way. 
This was done in \cite{LH2008}.   

More precisely, one has 
\eqb
\label{moddisplaw}
\omega^2=\vec{p}^2+G(|\vec{p}|,T)\,,
\eqe
where $\omega$ denotes the frequency, $\vec{p}$ spatial momentum, and 
the function $G$ can be positive (screening) or 
negative (screening). It was shown in \cite{LH2008} and also in \cite{KH2007} 
that within any experimentally feasible accuracy 
the selfconsistent determination of $G$ only involves the one-loop tadpole 
diagram\footnote{A potential imaginary contribution to $G$, arising 
from the other one-loop diagram, vanishes identically on the mass shell 
in Eq.\,(\ref{moddisplaw}) \cite{LH2008}.}.     

The implications of this result for the 
low-frequency ($\omega\le 0.15\,T$) and low temperature ($T_c\le T\le 5\,T_c$) regime 
of the associated black-body spectrum were discussed in 
\cite{SHG2006II} for $p^2=0$ and in \cite{LH2008} for $p^2=G$ 
after postulating that an SU(2) gauge principle underlies 
photon {\sl propagation}. For $T\le 2\,T_c$ the former and the later 
results coincide while at higher temperatures the selfconsistent 
calculation indicates a decay $\propto T^{-1/2}$ of the spectral gap 
as compared to the result $\propto T^{1/3}$ obtained 
by assuming $p^2=0$ \cite{SHG2006I}. 

To avoid a contradiction with precision measurements of the present CMB intensity 
observing a perfect U(1) (Planckian) spectrum 
(spectral deviations leading to $\delta T/T\sim 10^{-5}$ only), 
the Yang-Mills scale $\Lambda_{\tiny\mbox{CMB}}$ of this SU(2) 
theory must be such that the present CMB temperature $T_{\tiny\mbox{CMB}}\sim 2.73\,$K 
coincides with the critical temperature $T_c$ of the deconfining-preconfining phase 
transition \cite{H2009}. 

In the SU(2) based black-body spectrum, which exhibits its 
largest deviation from the Planckian spectrum at $T\sim 2\,T_c$, 
there is an exponentially with increasing frequency decaying regime of antiscreening (photon's energy less 
than the modulus of its spatial momentum). 
At low frequencies the modified black-body spectrum exhibits 
screening (photon's energy larger than the modulus of its spatial momentum) 
which is induced by the presence 
of static monopoles\footnote{W.r.t. the defining SU(2) Yang-Mills fields 
these monopoles are magnetic. They have a dual interpretation 
in the Standard Model, and thus are electrically charged w.r.t. U(1)$_{\tiny\mbox{em}}$.}. 
By integrating over the spectral 
intensity these two effects do not cancel completely, and 
one obtaines the following integrated black-body anomaly:    
\eqb
\label{deviu1s2}
\delta\rho=\int_{0}^{\infty}\mathrm{d}\omega\,I_{\mathrm{SU}(2)}-\int_{0}^{\infty}\mathrm{d}\omega\,I_{\mathrm{U}(1)}<0\,,
\eqe
where 
\eqb
I_{\mathrm{SU}(2)}(\omega)=I_{\mathrm{U}(1)}(\omega)\times
\frac{(\omega-\frac{1}{2}\frac{\mathrm{d}}{\mathrm{d}\omega}G)\sqrt{\omega^{2}-G}}
{\omega^{2}}\theta(\omega-\omega^{*})\,,
\eqe
and 
\eqb
I_{\mathrm{U}(1)}(\omega)=\frac{1}{\pi^{2}}\frac{\omega^{3}}{\exp[\frac{\omega}{T}]-1}\,.
\eqe
It was argued in \cite{SzopaHofmann2007} that temperature $T$ can be regarded a scalar 
field which, in a given background cosmology, can be described by the following action: 
\begin{equation}
\label{Tdeppa}
\sqrt{-g}\,\mathcal{L}_{\tiny\mbox{CMB}}=\left(\frac{\bar T_0}{\bar T}\right)^3
\left(k\,\pd_\mu\delta T\pd^\mu\delta T-\delta\rho(T)\right)\,,
\end{equation}
where $k$ is a coefficient in need of empirical determination, and $\bar T$ is 
the mean temperature at a given redshift $z$. 
Defining a function $\hat{\rho}(T,T_0)$ as
\eqb
\label{hatrho}
\delta\rho=T_0^2\,\hat{\rho}\,,
\eqe
varying the action associated with Eq.\,(\ref{Tdeppa}) w.r.t. $\delta T=T-\bar{T}$, 
and linearizing the resulting equation of motion yields:
\eqb
\label{eqmotion}
\partial_{\tilde{\mu}}\partial^{\tilde{\mu}}\delta T-\frac{3}{\bar T}\,\pd_\tau\bar{T}\,\pd_\tau\delta T+
\frac{\bar{T}_0^2}{k H_0^2}\left[\frac12\,\left.\frac{\mathrm{d}^2\hat{\rho}}{\mathrm{d} T^2}\right|_{T=\bar{T}}\,\delta T+\frac12\,
\left.\frac{\mathrm{d}\hat{\rho}}{\mathrm{d} T}\right|_{T=\bar{T}}\right]=0\,.
\eqe
To arrive at Eq.\,(\ref{eqmotion}), the coordinate transformation
\eqb
\label{ct}
\tilde{x}_0\equiv\tau=H_0\,t\,,\ \ \ \ \ \ \ \ \ \tilde{x}_i=\frac{\mathrm{d}a}{\mathrm{d}t}\,x_i\,,\ \ \ \ \ (i=1,2,3)\,
\eqe
was performed. 
Notice the extremely large factor $(\bar{T}_0/H_0)^2\sim 10^{60}$ in front 
of the square brackets in Eq.\,(\ref{eqmotion}). This factor arises because we chose to 
measure time $\tau$ in units of the age of the Universe, distances from the origin 
$\tilde{x}_i$ in units 
of the actual horizon size $H^{-1}=a/\frac{\mathrm{d}a}{\mathrm{d}t}$ 
(as long as $|\tilde{x}_i|$ is sufficiently smaller than unity), and 
temperature in units of $\bar{T}_0=2.35\times 10^{-4}\,$eV.   

By assuming spherical symmetry for the fluctuation $\delta T$, which is relevant for an 
analysis of the cosmic dipole, Eq.\,(\ref{eqmotion}) simplifies as:
\eab
\label{difeqO}
0&=&\partial_{\tau}\partial_{\tau}\delta T-\left(\frac{\mathrm{d}a}{a\,\mathrm{d}\tau}\right)^2\,
\left[\partial_{\sigma}\partial_{\sigma}\delta T+\frac{2}{\sigma}\,\partial_{\sigma}\delta T\right]
-\frac{3}{\bar T}\,\pd_\tau\bar{T}\,\pd_\tau\delta T+\nonumber\\ 
&&\frac{\bar{T}_0^2}{k H_0^2}\left[\frac12\,\left.\frac{\mathrm{d}^2\hat{\rho}}{\mathrm{d} T^2}\right|_{T=\bar{T}}\,\delta T+\frac12\,
\left.\frac{\mathrm{d}\hat{\rho}}{\mathrm{d} T}\right|_{T=\bar{T}}\right]\,.
\eae
In Eq.\,(\ref{difeqO}) we have introduced 
$\sigma\equiv\sqrt{\tilde{x}_1^2+\tilde{x}_2^2+\tilde{x}_3^2}$. 
Erroneously, a factor $1/a^2$, $a=\frac{\bar{T}_0}{\bar{T}}$ 
being the scale factor normalized to unity today, 
was missing in Eq.\,(18) of \cite{SzopaHofmann2007} in the prefactor $-H^2/H_0^2=-
\left(\frac{\mathrm{d}a}{a\,\mathrm{d}\tau}\right)^2$ of the term containing spatial derivatives,  
and the function $\hat{\rho}$ was used as obtained in 
the approximation $p^2=0$. The main goal of the present article 
is to investigate to what extent these unaccuracies affect 
the result $\delta T$ in the spatially spherically symmetric case.    

Eq.\,(\ref{difeqO}) is a two-dimen\-sio\-nal wave equa\-tion with
additional terms arising on one hand due to the time-dependence of the cosmological
background ($-\frac{3}{\bar T}\,\pd_\tau\bar{T}\,\pd_\tau\delta T$) and
on the other hand due to the presence of the black-body anomaly: The
term $\frac12
\frac{\bar{T}_0^2}{k
  H_0^2}\left.\frac{\mathrm{d}^2\hat{\rho}}{\mathrm{d}
    T^2}\right|_{T=\bar{T}}\,\delta T$ will be referred to as `restoring term',
and the term $\frac12\,\frac{\bar{T}_0^2}{k
  H_0^2}\left.\frac{\mathrm{d}\hat{\rho}}{\mathrm{d}T}\right|_{T=\bar{T}}$ will be 
referred to as `source term' in the following.    

\section{Old versus corrected results\label{OVN}}

The background cosmology used to evolve $\delta T$ according to 
Eq.\,(\ref{difeqO}) is the same spatially flat $\Lambda$CDM 
model that was used in \cite{SzopaHofmann2007}. 

\begin{figure}
\begin{center}
\leavevmode
\leavevmode
\vspace{6.3cm}
\includegraphics{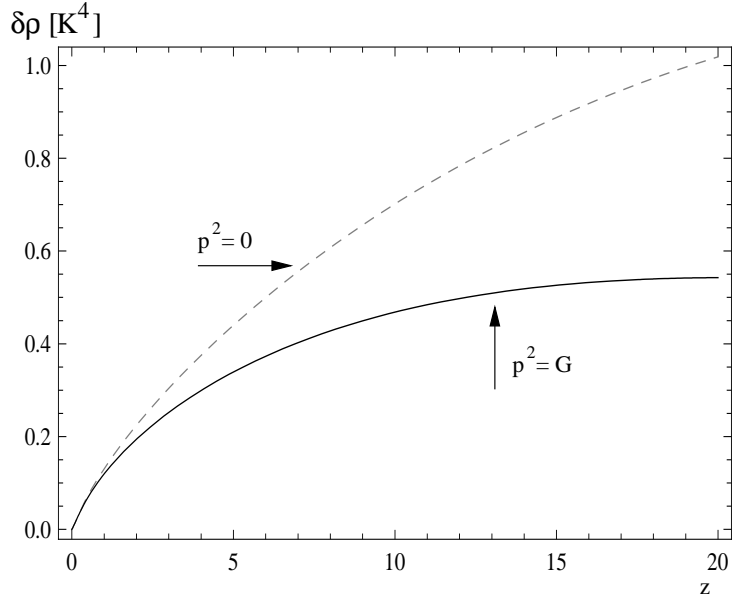}
\end{center}
\caption{The difference $\delta\rho$ between integrated modified black-body spectral intensity 
and the integrated Planckian spectrum as a function of redshift $z$. 
The approximation $p^2=0$ \cite{SzopaHofmann2007} corresponds to the dashed line while the case of the 
selfconsistently determined mass shell \cite{LH2008} is depicted by 
the solid line. \label{Fig-0}}      
\end{figure}

\begin{figure}
\begin{center}
\leavevmode
\leavevmode
\vspace{6.3cm}
\includegraphics{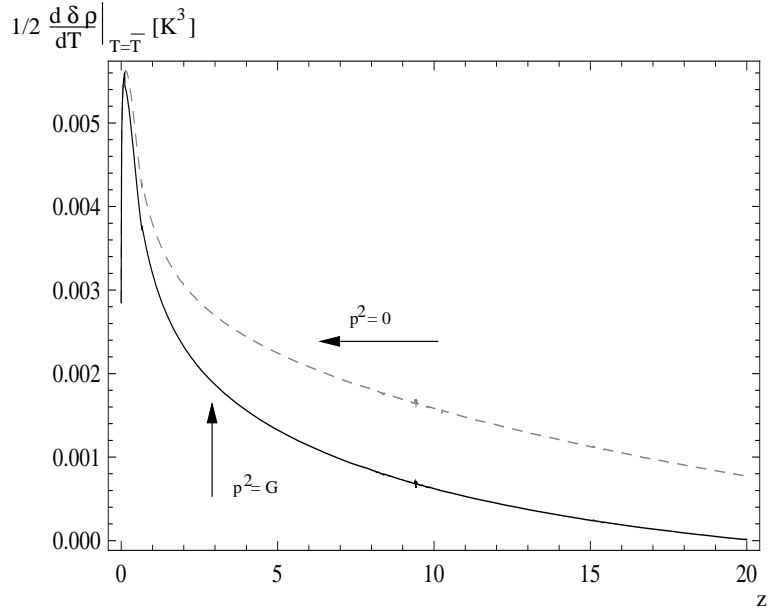}
\end{center}
\caption{The `source term'
  $\frac12\left.\frac{\mathrm{d}\,\delta\rho}{\mathrm{d}
      T}\right|_{T=\bar{T}}$ in Eq.\,(\ref{difeqO}) as a 
function of redshift $z$. The approximation $p^2=0$ \cite{SzopaHofmann2007} corresponds to the dashed 
line while the case of the selfconsistently determined mass shell \cite{LH2008} is depicted by the 
solid line. \label{Fig-1}}      
\end{figure}

\begin{figure}
\begin{center}
\leavevmode
\leavevmode
\vspace{4.5cm}
\includegraphics{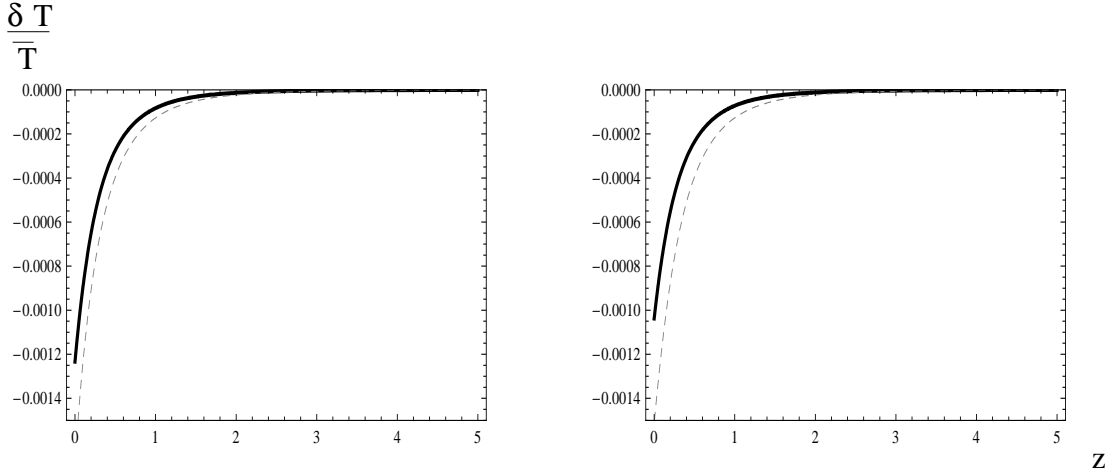}
\end{center}
\caption{$\frac{\delta T}{\bar T}$ for two distances $\sigma=0.05$ (left panel) and $\sigma=0.5$ 
(right panel) as a function of $z$ for $k=0.01868\,{\bar T}_0^2/H_0^2$. The width $w$ 
of the initial Gaussian is $w=10^{-2}$ at $z_i=20$. 
The approximation $p^2=0$ \cite{SzopaHofmann2007} together with the erroneous version 
of Eq.\,(\ref{difeqO}) corresponds to the dashed line 
while the case of the selfconsistently determined mass shell \cite{LH2008} together with the 
proper evolution equation Eq.\,(\ref{difeqO}) is depicted by the 
solid line.\label{Fig-2}}      
\end{figure}

\begin{figure}
\begin{center}
\leavevmode
\leavevmode
\vspace{6.3cm}
\includegraphics{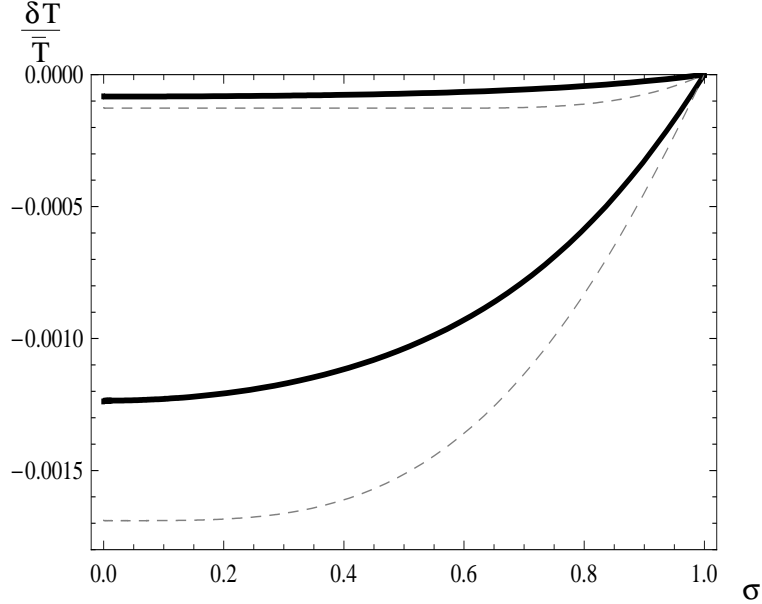}
\end{center}
\caption{The profile $\frac{\delta T}{\bar T}$ at $z=1$ (upper curves) and $z=0$ 
(lower curves). The solid lines correspond the case of the selfconsistently determined mass shell \cite{LH2008} together with the 
proper evolution equation (\ref{difeqO}) while the dashed lines are 
for the approximation $p^2=0$ \cite{SzopaHofmann2007} together with the erroneous version 
of Eq.\,(\ref{difeqO}). Again, we have used $k=0.01868\,{\bar T}_0^2/H_0^2$.
\label{Fig-3}}      
\end{figure}

\begin{figure}
\begin{center}
\leavevmode
\leavevmode
\vspace{6.3cm}
\includegraphics{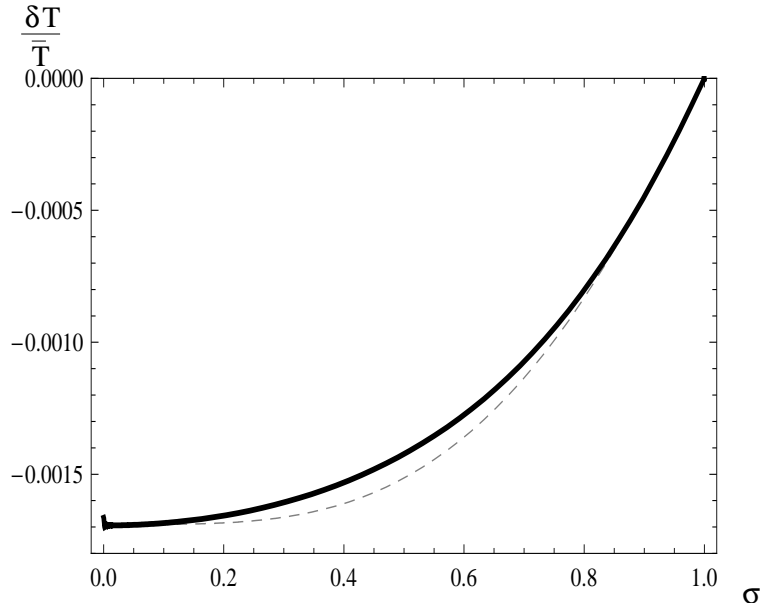}
\end{center}
\caption{The profile $\frac{\delta T}{\bar T}$ at $z=0$. The dashed line is 
for the approximation $p^2=0$ \cite{SzopaHofmann2007} together with the erroneous version 
of Eq.\,(\ref{difeqO}) and $k=0.01868\,{\bar T}_0^2/H_0^2$. 
The solid line is obtained for the case of the selfconsistently determined mass shell 
\cite{LH2008} together with the 
proper evolution equation (\ref{difeqO}) and $k=0.0136\,{\bar T}_0^2/H_0^2$. This 
$k$-value is generated by demanding that the function $\frac{\delta T}{\bar T}$ 
coincides with the dashed line at $\sigma=0$. 
\label{Fig-4}}      
\end{figure}

In Fig.\,\ref{Fig-0} we plot $\delta\rho$ as a function of redshift 
$z=a^{-1}-1$ as obtained in the approximation $p^2=0$ \cite{SzopaHofmann2007} (dashed line) 
and selfconsistently \cite{LH2008} (solid line). 
In Fig.\,\ref{Fig-1} we plot 
$\left.\frac{d\delta\rho}{dT}\right|_{T=\bar{T}}$ as a function of redshift 
as obtained in the approximation $p^2=0$ \cite{SzopaHofmann2007} (dashed line) 
and selfconsistently \cite{LH2008} (solid line). 
There is hardly any visually discernable difference between the results for the 
`restoring term' in Eq.\,(\ref{difeqO}) as obtained in the approximation $p^2=0$ \cite{SzopaHofmann2007} 
and selfconsistently \cite{LH2008}. In Fig.\,\ref{Fig-2} a plot of 
$\frac{\delta T}{\bar T}$ is shown as a function of $z$ for fixed distances 
$\sigma=0.05;0.5$, $k=0.01868\,{\bar T}_0^2/H_0^2$, and a width $w$ 
of the initial Gaussian of $w=10^{-2}$ at $z_i=20$. 
The approximation $p^2=0$ \cite{SzopaHofmann2007} together with the erroneous version 
of Eq.\,(\ref{difeqO}) corresponds to the dashed line 
while the case of the selfconsistently determined photon mass shell \cite{LH2008} together with the 
proper evolution equation Eq.\,(\ref{difeqO}) is depicted by the solid line.
In Fig.\,\ref{Fig-3} plots of the profiles $\frac{\delta T}{\bar T}$ at $z=1$ and $z=0$ are 
shown when computed both with the erroneous version of Eq.\,(\ref{difeqO}) together with the 
approximation $p^2=0$ and with the proper equation (\ref{difeqO}) and 
the case of the selfconsistently determined photon mass shell.  
In Fig.\,\ref{Fig-4} plots of the profiles $\frac{\delta T}{\bar T}$ at $z=0$ are 
shown when computed both with the erroneous version of Eq.\,(\ref{difeqO}) together with the 
approximation $p^2=0$ and with the proper equation (\ref{difeqO}) and 
the case of the selfconsistently determined mass shell. Notice that 
$k=0.01868\,{\bar T}_0^2/H_0^2$ in the former case while the $k$-value for the latter case was adjusted 
such that the two profiles coincide at $\sigma=0$. Considering the empirical uncertainty of 
$k$ \cite{SzopaHofmann2007} this adjustment is admissible. Obviously, the difference between 
the two curves is small.

\section{Summary and conclusions\label{SC}}

In this work we have investigated to what extent a calculational error in the derivation of the evolution equation for 
temperature fluctuations $\delta T$ and an improvement \cite{LH2008} in determining the black-body-anomaly 
related terms in this equations affects the results for the evolution of temperature profiles in the 
spatially spherically symmetric case \cite{SzopaHofmann2007}. 
We have found that only very small quantitative deviations take place which 
practically can be absorbed in the empirical 
uncertainty of the normalization of the kinetic 
versus the potential term in the associated action (\ref{Tdeppa}). 

To make contact with the observed large-angle anomalies in the CMB 
it would be interesting to compute the dipole-subtracted angular correlation function 
$C(\theta)=\la \delta T\delta T\ra$. In the approach discussed in \cite{SzopaHofmann2007} 
and again applied here this would be done by 
relaxing the assumption of spherical symmetry made in Eq.\,(\ref{difeqO}) to 
compute an ensemble of present CMB maps of temperature fluctuations (solutions to Eq.\,(\ref{eqmotion}) minus solutions to 
Eq.\,(\ref{difeqO}), labelled by the initial conditions and by 
subsequently averaging over this ensemble and over pairs of integrals over 
lines of sight separated by the angle $\theta$. Due to the rapid build-up of the spherical 
profile at $z\sim 1$ we expect an anomalous suppression of large-angle correlations 
and a statistically significant alignment of the low multipoles.

\end{document}